# Cultural values and cross-cultural video consumption on YouTube

**Minsu Park[1]¶, Jaram Park[2]¶, Young Min Baek[3]\*, and Michael Macy[1,4]\***

[1] Department of Information Science, Cornell University, Ithaca, USA, 14850

[2] Graduate School of Culture Technology, KAIST, Daejeon, Republic of Korea, 305-701

[3] Department of Communication, Yonsei University, Seoul, Republic of Korea, 120-749

[4] Department of Sociology, Cornell University, Ithaca, USA, 14850

\* Corresponding author

Email: ymbaek@yonsei.ac.kr (YMB) or mwm14@cornell.edu (MM)

¶ These authors equally contributed to this work.

# Abstract

Video-sharing social media like YouTube provide access to diverse cultural products from all over the world, making it possible to test theories that the Web facilitates global cultural convergence. Drawing on a daily listing of YouTube's most popular videos across 58 countries, we investigate the consumption of popular videos in countries that differ in cultural values, language, gross domestic product, and Internet penetration rate. Although online social media facilitate global access to cultural products, we find this technological capability does not result in universal cultural convergence. Instead, consumption of popular videos in culturally different countries appears to be constrained by cultural values. Cross-cultural convergence is more advanced in cosmopolitan countries with cultural values that favor individualism and power inequality.

# Introduction

The recent upsurge of nationalist movements opposing open borders and free trade brings new urgency to questions about the effects of social media on cultural convergence. Video-sharing social media like YouTube provide access to diverse cultural products from all over the world [1]. Unlike traditional media such as television, CDs, or books [2], content on social media (e.g., video clips and music videos) is readily accessible across countries that differ in national GDP [3,4], geographic location [3,5], language [6], and religion [5].

Nevertheless, the ability to easily obtain social media content does not mean consumers take advantage of the opportunity. Although technologies increasingly facilitate cross-border flow of media content, previous studies support the "cultural proximity hypothesis" that consumption reflects cultural values that in turn shape cultural norms about socially acceptable content, such that consumers prefer products closer to their own culture [4,7–9]. However, these studies focused on consumption of tangible cultural products like books and CDs [7], not content on social media that can be easily downloaded from the Web. Hyperlinks on web pages [4,8] were also studied extensively, but hyperlinks are generated by producers of online content who vie for the attention of the public and hyperlinks themselves do not reveal consumption patterns of online content.

An important exception is a study [1] showing that Korean pop (or K-pop) music videos on YouTube are highly popular in countries whose cultures differ sharply from Korea as well as in countries that are culturally very similar to Korea. However, it remains to be seen whether this finding generalizes beyond one type of media content produced in only one country.

Using co-consumption of popular videos on YouTube, this study extends research on the cultural proximity hypothesis by examining the relationship between cultural values and cultural openness. Drawing on a daily listing of YouTube's most popular videos across 58 countries, we investigate the consumption of popular videos in countries that differ in cultural values [10], language, gross domestic product (GDP), and Internet penetration rate.

We chose YouTube because it is the most popular platform for media consumption on the Web, with more than one billion viewers every day, watching hundreds of millions of hours of content [11,12]. Video over Internet Protocol is forecast to represent 82 percent of all download traffic by 2020 [13]. Our research addresses why some YouTube videos (e.g., Gangnam Style) are globally consumed while others are limited to a single country, despite the existence of a technological infrastructure for global cross-cultural communication. To find out, we recorded the 50 most popular videos listed by YouTube for the past day for each of 74 countries over six months. "Popularity" is based on YouTube's undisclosed algorithm [3,14] that takes into account views, downloads, and likes. Inclusion on YouTube's top 50 list provides an unranked measure of video consumption.

## Cultural values

"Culture" has been defined as a set of values maintained across generations through the socialization process [10,15]. Although individual attitudes and beliefs may be in constant flux, cultural values are thought to be stable attributes of societies [16]. Cultural values are defined as enduring beliefs that "a specific mode of conduct or end-state of existence is personally or socially preferable" [9, p.5]. These cultural values influence user decisions about what to view, download, or like, which suggests that YouTube video consumption can be expected to vary across cultures [17,18].

We operationalized cultural values using Hofstede's four-dimensional model [10], based on aggregated survey responses from IBM employees in 76 countries. Hofstede's approach has been criticized by culture scholars who argue that culture is too subtle to quantify, especially in multi-cultural countries like the United States [19,20]. Nevertheless, Hofstede's measures have been widely applied in prominent studies showing how cultural values influence cross-cultural communication behaviors such as media selection and adoption [1,21], political discussion engagement [22], and use of emoticons on Twitter [23].

The four dimensions in Hofstede's model are *individualism (IDV)*, *uncertainty avoidance (UAI)*, *power distance (PDI)*, and *masculinity (MAS)*. Each of these dimensions has implications for cross-cultural media consumption that we operationalize in turn below.

## *Individualism-collectivism (IDV)*

Countries with high IDV are more inclined to emphasize "I" rather than "we" and to privilege individual interests over collective welfare [10, p.130]. Individualistic cultures do not demand conformity around shared opinions, beliefs, or attitudes and are therefore more likely to embrace cultural diversity and to show "respect for other cultures" [10, p.99]. Hofstede's argument has been supported by studies [16] showing that people in individualistic cultures tend to be more tolerant of diversity and appreciative of cultural differences. Other studies have found that people in high IDV countries consume more cross-national products [24], adopt global platforms like B2C e-commerce [25] and SNS [21], and purchase newly launched brands [26]. Using a large international hyperlink network, Barnett and Sung [4] found that high IDV countries occupied more central positions in the international information-sharing network. We hypothesize that this pattern will extend to cross-cultural video consumption on YouTube:

H1: People in individualistic countries will be more likely to consume videos that are popular in culturally different countries, compared to those in collectivistic countries.

## *Uncertainty avoidance index (UAI)*

People in high UAI countries are more likely to "feel threatened by ambiguous or unknown situations" [10, p.191]. For example, opinion surveys have found that people in European countries with high UAI scored higher in aggressive nationalism, ethnocentrism, and xenophobia, including beliefs that immigrants should be sent back to their countries of origin [27]. People living in high UAI countries are reluctant to purchase newly launched products or adopt technological innovations, including the Internet [28], mobile phones [29], SNS [30], and B2C e-commerce [25]. This attitude may extend to consumption of foreign videos:

H2: Countries with high uncertainty avoidance will be less likely to consume videos that are popular in culturally different countries, compared to those in low uncertainty avoidance countries.

## *Power distance index (PDI)*

People in high PDI countries are more likely to "expect and accept that power is distributed unequally" [10, p.61] in groups or organizations. Because Hofstede's measure is based on survey responses of employees, PDI applies most directly to the relationship between bosses and subordinates in organizations [10]. Thus, few studies have examined how PDI influences cross-cultural behavior. However, PDI has implications for beliefs about status inequality that imply cultural preferences for products that signal cultural superiority and it has been demonstrated that people in cultures with high PDI tend to consume products that help them establish and express their status [31]. Bourdieu's classic study [32] shows how cultural products are used to construct and define social class hierarchies. People with high status are believed to have more cultural sophistication, including more extensive and detailed knowledge about foreign cultures. Foreign products provide symbolic benefits such as modernity, prestige, and associations with foreign lifestyles [33] in a similar manner that products with recognized, exclusive, and relatively expensive brand names tend to have higher levels of social status attached to them compared to more generic and less exclusive brands [26]. In high PDI countries, these symbolic benefits constitute a primary motivation for foreign product consumption [33]. This suggests the possibility that people in high PDI countries (including elites as well as those with elite pretensions) tagare more likely to regard xenophilia as a signal for cultural sophistication [26,31]. More formally, we expect:

H3:  People in high PDI countries will be more likely to consume videos that are popular in culturally different countries than those in low PDI countries.

## *Masculinity (MAS)*

People in high MAS countries are more likely to conform to gender role stereotypes that "men are supposed to be assertive, tough, and focused on material success, whereas women are supposed to be more modest, tender, and concerned with quality of life" [10, p.140]. Even more than with PDI, MAS does not have straightforward implications for cross-cultural media consumption. On the one hand, it might be argued that masculinity encourages cultural boldness, which implies a greater likelihood to consume unfamiliar cultural content. On the other, traditional gender roles may be associated with parochial cultural tendencies, which implies the opposite association. Moreover, these opposing effects may cancel each other out. We therefore do not hypothesize an association in either direction but instead test to see if high and low MAS countries differ in video consumption.

# Materials and methods

## *YouTube data collection*

YouTube only provides aggregate country-level measures of popularity and we therefore do not have individual user-level data. For each country, YouTube lists daily the "most popular" videos, accessible through the YouTube Application Programming Interface (API). We collected the 50 most popular videos for each of 74 countries over 6 months from November 15th 2014 to April 5th 2015 (approximately 40,700 observations per day) [34] for a total of 4,979,077 observations and 561,931 unique videos. Each observation contains the date, category, title, tags, video duration, average view duration, comments, and popularity metrics, including the number of views, likes, dislikes, and shares for that day.

## *Bipartite co-consumption network*

We used the pairwise co-listings of popular videos to construct a bipartite projected network of countries. We first built a bipartite network as proposed by [3,6] with two types of nodes: (1) 74 countries, each with a list of popular videos collected from YouTube and (2) 561,931 videos on those countries' popular video lists. In the projection of this bipartite network, each country was regarded as a node and an edge was assigned if a pair of countries shared one or more videos on their popular video lists. As it happens, all countries were connected, that is, all had at least one overlapping video with another country. Following the method suggested by Newman [35], an edge weighting was applied that privileges videos that appear less frequently across all 74 lists. Thus, a pair of countries that co-lists a set of videos that are universally popular has relatively low weight compared to a pair of countries that has in common a set of videos that appear on no other lists. This weighting method mitigates the effect of overly popular outliers (potentially due to an artificial increase of viewers) on the co-consumption patterns. This edge weighting combines two components: (1) the number of videos co-listed by a pair of countries and (2) the global popularity (or out-degree) of each video in the co-list, defined as the number of countries in which the video was listed. The weighting metric was formalized as follows:

$$W_{ij} = \sum_{k} \frac{\delta_i^k \delta_j^k}{n_k - 1}$$

where $W_{ij}$ is the weight of the edge between countries $i$ and $j$; $k$ is a unique YouTube video in the set of videos co-listed by $i$ and $j$; $n_k$ is the number of countries that listed $k$; and $\delta_i^k$ is 1 if video $k$ is co-listed on popular video lists including country $i$ and 0 otherwise.

Fig 1A illustrates the co-consumption pattern on a stylized bipartite network, and 1B shows how the edge weight is computed between a pair of countries, the US and Germany. The first video is popular only in the U.S. and Germany, giving it a weight of 1. The weight of the last video is 1/4 because it appears on the most popular lists of five different countries (including the U.S. and Germany). We then derive the edge weight as the sum the weights over all the co-listed videos. Thus, the edge weight reflects the number of co-listed videos weighted by the inverse of the video out-degree (the number of countries that list that video). In Fig 1, the edge weight between the U.S. and Germany is 7/4. Using this weighting method, we computed edge weights between all possible pairs of countries.

**Fig 1. Construction of the bipartite network of video co-consumption on YouTube.**

[Fig 1 here]

The outcome of the final step is shown in Fig 1C, in which the edge weights are filtered to preserve only those edges that deviate from the expected weight in a null model iteratively produced by a random assignment from a uniform distribution. By imposing a significance level of *p < 0.05*, the links whose weights exceed a randomly expected value are preserved. The remaining links constitute the "backbone" structure of the network [36].

## *Measure of cultural openness*

We refer to "cultural openness" as the conceptual outcome of interest. A country with high cultural openness can achieve an "optimal blend of novelty and familiarity" by creating cultural bridges [37, p.11824]. At the same time, a country with high cultural openness can co-consume cultural products with many other countries across different cultural clusters such that they are close to most other countries in the network, which can be perceived as "openness to diversity." Instead of using a single measure that combines these aspects, we operationalize cultural openness as having two distinct dimensions. Betweenness centrality was used as an indicator of bridging between cultures, measured as video overlap with other countries that do not overlap with one another. Closeness centrality, measured as a country's level of video overlap with all other countries, provides an indicator of cultural diversity.

We measured cultural betweenness and closeness using Opsahl et al.'s [38] centrality measures in weighted networks to take both the number of ties and the tie weights into account. Those weighted centrality measures are variants of Djikstra's algorithm, a well-known method for finding and computing the shortest paths among nodes in a network. Using this approach, the shortest path *d* between two nodes *(i, j)* can be defined as follows:

$$d^{w\alpha}(i,j) = \min(\frac{1}{(w_{ih})^\alpha} + ... + \frac{1}{(w_{hj})^\alpha})$$

where *w* is the weight of the tie between nodes; *h* are intermediary nodes on paths between node *i* and *j;* and $\alpha$ is a tuning parameter that reflects the influence of edge weights. When $\alpha$ = 0, Opsahl's algorithm reduces to the familiar binary measure in which a network edge either exists or does not (i.e., the level of similarity or affiliation between countries can not be captured at all because tie weights are ignored). When $\alpha$ = 1, the algorithm is identical to Dijkstra's (i.e., the original feature of the measures, particularly the number of ties, is ignored because tie weights are the sole determinant). A value for $\alpha$ < 1 assigns the path with the greatest number of intermediary nodes the longest distance whereas the impact of additional intermediary nodes is relatively unimportant compared to the strength of the ties when $\alpha$ > 1. Hence, for $\alpha$ < 1, a shorter path composed of weak ties is favored over a longer path with strong ties. Conversely, for $\alpha$ > 1, paths with more intermediaries connected by strong ties are favored. The tuning parameter is used to operationalize the extent to which openness reflects a more balanced weight distribution in a node's local network along with its degree. We set $\alpha$ = 0.5, although results are fairly robust across other values of the tuning parameter smaller than or equal to 1. Formally, cultural betweenness is given by:

$$C_B^{w\alpha}(i) = \frac{g_{jk}^{w\alpha}(i)}{g_{jk}^{w\alpha}}$$

where *g* is the sum of shortest paths that pass through node *i* as a proportion of all shortest paths in the network. Cultural closeness, as the inverse sum of shortest distances to all other nodes from a focal node, is given by:

$$C_C^{w\alpha}(i) = \left[\sum_{j=1}^{N} d^{w\alpha}(i,j)\right]^{-1}$$

We limited the analysis to the 58 countries for which we had Hofstede scores. The list of countries included and excluded in the analysis is provided in S1 Table. Descriptive statistics of the four scores on the 58 countries are: (1) individualism ($M = 41.00$, $SD = 23.13$), (2) uncertainty avoidance ($M = 66.72$, $SD = 22.81$), (3) power distance ($M = 61.95$, $SD = 21.59$), and (4) masculinity ($M = 49.55$, $SD = 17.01$).

### *Economic, linguistic, and technological measures*

To disentangle cultural influence from other factors that have been found to affect cultural openness, we included economic, linguistic, and technological measures. Per capita GDP has been shown to be strongly associated with cross-cultural communication on Twitter [39] and in international transactions and communication flows [4,40]. Previous research also shows strong correlations between GDP per capita and Hofstede's cultural values [10,41]. In short, GDP per capita is associated with both cultural openness and cultural values. We used the GDP per capita data archived by the World Bank in 2013 [42]. Since the average GDP per capita across 58 countries showed a right-skewed distribution, the base 10 log-transformed GDP per capita ($M = 4.06$, $SD = .56$, $Median = 4.17$) was used in the analysis.

Language is an obvious barrier to any global communication [7,8] and social media interaction in particular [6,43]. As a consequence, English as a *lingua franca* allows greater access to cultural diversity [6,8], compared to local languages such as Korean or Japanese. Following Ronen et al.'s [6] algorithm, we computed eigenvector centrality of a country's language. The average eigenvector centrality of language across 58 countries is $M = 0.18$ ($SD = .32$, $Median = .025$). The higher the centrality, the lower the linguistic barriers to global communication.

Internet penetration is strongly correlated with Hofstede's cultural values [28] and also limits access to online cultural content. We used the World Bank measure of Internet penetration as the number of Internet users per 100 people [42]. The distribution of Internet penetration is normally distributed ($M = 64.90$, $SD = 21.67$, $Median = 69.48$, ranging from 12.30 to 95.05) and thus does not require log transformation.

## Results

Table 1 reports descriptive statistics for the video co-consumption networks across different categories that were automatically classified by YouTube. Each network is based on the edge weights derived from the co-listing of videos in a particular category. The number of nodes is not identical across categories because of data sparsity. Countries were deleted for which there were too few videos listed in that category to obtain statistically significant links in the "backbone" network.

**Table 1. Descriptive statistics of network structure by video category.**

| Category | Nodes | Edges | Degree | Weighted Degree | Modularity | CC | APL |
|---|---|---|---|---|---|---|---|
| Combined | 72 | 195 | 5.417 | 324.864 | 0.736 | 2 | 3.067 |
| News | 73 | 263 | 7.205 | 628.543 | 0.667 | 2 | 2.887 |
| Music | 68 | 159 | 4.676 | 145.903 | 0.410 | 3 | 3.332 |
| Games | 72 | 168 | 4.667 | 574.939 | 0.767 | 5 | 3.812 |
| Sports | 70 | 173 | 4.943 | 391.373 | 0.695 | 3 | 4.406 |
| Entertainment | 72 | 217 | 6.028 | 461.752 | 0.675 | 4 | 3.036 |
| Film | 73 | 205 | 5.616 | 401.584 | 0.637 | 1 | 3.480 |
| People | 73 | 257 | 7.041 | 469.417 | 0.624 | 1 | 2.917 |
| Tech | 68 | 179 | 5.265 | 293.704 | 0.619 | 2 | 3.383 |
| Comedy | 72 | 179 | 4.972 | 324.729 | 0.619 | 2 | 3.232 |
| Travel | 72 | 228 | 6.333 | 300.295 | 0.598 | 2 | 2.857 |

Note: CC = Connected Components; APL = Average Path Length

These network characteristics reveal interesting differences across video categories. For example, the co-consumption news network has low average path length (APL) but high average degree, indicating that news videos were more likely to be consumed among a broader global audience than other types of videos. In contrast, the co-consumption music network has fewer nodes and edges and has low average degree and high APL, which indicates that a country's video list contained less globally popular music and more locally popular music as described in [44]. Interestingly, the gaming network has high modularity and many connected components (CCs), indicating more clustered video preferences.

Although these patterns invite category-specific analyses, the theoretical motivation for this study is focused on differences between countries, not differences between cultural categories. We therefore report results for the combined network based on all videos, regardless of category. However, we also checked the robustness of the overall pattern by examining each category-specific network and found no important differences.

## *The relationships between cultural values and cultural openness*

Tables 2 and 3 report results for regression analyses of cultural openness, operationalized as cultural betweenness and closeness. The cultural model consists of Hofstede's four cultural values and the non-cultural model includes economic, linguistic, and technological measures. The combined model reports the effects of cultural values net of non-cultural.

**Table 2. OLS regression model of cultural betweenness among 58 countries.**

|  | **Full model** | **Non-culture Model** | **Culture Model** |
|---|---|---|---|
| **Intercept** | -0.425* | 0.014 | -0.339* |
|  | (0.172) | (0.073) | (0.147) |
| **Non-cultural factors** |  |  |  |
|   GDP per capita (log-transformed) | 0.092 | 0.087 |  |
|  | (0.219) | (0.228) |  |
|   Language eigenvector centrality | 0.092 | 0.173* |  |
|  | (0.085) | (0.077) |  |
|   Number of Internet users | 0.028 | 0.026 |  |
|  | (0.240) | (0.242) |  |
| **Cultural values** |  |  |  |
|   Individualism (IDV) | 0.354* |  | 0.426** |
|  | (0.138) |  | (0.123) |
|   Uncertainty avoidance (UAI) | -0.031 |  | -0.061 |
|  | (0.122) |  | (0.113) |
|   Power distance (PDI) | 0.410* |  | 0.373* |
|  | (0.156) |  | (0.147) |
|   Masculinity (MAS) | 0.214 |  | 0.250* |
|  | (0.125) |  | (0.124) |
| **Sample size (number of countries)** | 58 | 58 | 58 |
| **Model-fit indices** |  |  |  |
|   $R^2$ | 0.290 | 0.110 | 0.263 |
|   Adjusted $R^2$ | 0.191 | 0.061 | 0.208 |

Note: * $p < .05$, ** $p < .01$, *** $p < .001$. Unstandardized coefficients are reported with standard errors in parentheses. In order to compare coefficients, variables included in the analyses were rescaled to the unit interval.

**Table 3. OLS regression model of cultural closeness among 58 countries.**

|  | Full model | Non-culture Model | Culture Model |
|---|---|---|---|
| **Intercept** | 0.138 | 0.368*** | 0.285** |
|  | (0.109) | (0.051) | (0.100) |
| **Non-cultural factors** |  |  |  |
| GDP per capita (log-transformed) | 0.206 | 0.193 |  |
|  | (0.139) | (0.161) |  |
| Language eigenvector centrality | 0.074 | 0.190** |  |
|  | (0.54) | (0.055) |  |
| Number of Internet users | 0.014 | 0.001 |  |
|  | (0.153) | (0.171) |  |
| **Cultural values** |  |  |  |
| Individualism (IDV) | 0.229* |  | 0.328*** |
|  | (0.088) |  | (0.084) |
| Uncertainty avoidance (UAI) | -0.229** |  | -0.234** |
|  | (0.077) |  | (0.077) |
| Power distance (PDI) | 0.314** |  | 0.236* |
|  | (0.099) |  | (0.100) |
| Masculinity (MAS) | 0.245** |  | 0.270** |
|  | (0.086) |  | (0.084) |
| **Sample size (number of countries)** | 58 | 58 | 58 |
| **Model-fit indices** |  |  |  |
| $R^2$ | 0.540 | 0.284 | 0.452 |
| Adjusted $R^2$ | 0.476 | 0.244 | 0.411 |

Note: * $p < .05$, ** $p < .01$, *** $p < .001$. Unstandardized coefficients are reported with standard errors in parentheses. In order to compare coefficients, variables included in the analyses were rescaled to the unit interval.

We tested both models for heteroscedasticity and multicollinearity. For cultural closeness, we could not reject the null hypothesis that the variance of the residuals is constant, i.e. heteroscedasticity is not present, using the test of non-constant variance score. For the model of cultural betweenness, in contrast, we inferred that the residuals are heteroscedastic. However, as described earlier, results are robust across different tuning parameters and a model constructed with a composite measure of betweenness and closeness [45,46] in S1 Text also shows qualitatively similar results to cultural betweenness and closeness models. The variance inflation factor on each variable of the full model is smaller than two except for GDP per capita (2.41) and Internet diffusion (2.49) that are strongly correlated with each other but neither contributes significantly to model predictions.

As shown in Tables 2 and 3, a country's cultural openness is much better explained by cultural values (adjusted $R^2$ = .208 for cultural betweenness; adjusted $R^2$ = .411 for cultural closeness) than non-cultural measures (adjusted $R^2$ = .061 for cultural betweenness; adjusted $R^2$ = .244 for cultural closeness). Indeed, non-cultural measures do not make a significant contribution to the full models' explanatory power, and removing these measures even improves the adjusted $R^2$ (.191) of the model on cultural betweenness.

The coefficients in Tables 2 and 3 provide more detailed results. The eigenvector centrality of language [6] indicates that countries using more global languages (e.g., English) have greater cultural openness, consistent with the findings in previous studies ($b$ = .173, $p$ < .05 for cultural betweenness; $b$ = .190, $p$ < .01 for cultural closeness) [7,8]. However, this effect largely disappears when cultural values are included in the model ($b$ = .092, $p$ = .29 for cultural betweenness; $b$ = .075, $p$ = .18 for cultural closeness), indicating that cultural values are stronger predictors of cultural openness and capture most of the linguistic effect.

The results in Table 2 (cultural betweenness) support H1 and H3 but not H2. As hypothesized, YouTube users consume more videos in common with other countries that do not overlap with one another if those users are located in countries that are more individualistic ($b$ = .354, $p$ < .05) and with greater power distance ($b$ = .410, $p$ <

.05). However, a country's cultural openness is not predicted by uncertainty avoidance ($b$ = -.031, $p$ = .80) or masculinity ($b$ = .214, $p$ = .12). In short, individualism and acceptance of power inequality are associated with an optimal blend of novelty and familiarity, as indicated by greater cultural betweenness.

Results in Table 3 (cultural closeness) support H1, H2, and H3. YouTube users consume more videos in common with a larger number of culturally diverse countries (i.e., higher cultural closeness) if those users are located in countries that are more individualistic ($b$ = .229, $p$ < .05), with less uncertainty avoidance ($b$ = −.229, $p$ < .01), greater power distance ($b$ = .314, $p$ < .01), and higher conformity to gender role stereotypes ($b$ = .245, $p$ < .01). In short, cultural closeness is associated with more cultural values than is cultural betweenness, and both are more important than the non-cultural factors that have been the focus of previous research.

## Discussion

Our findings are consistent with the view that cross-cultural convergence, especially cultural closeness, is more advanced in cosmopolitan countries with cultural values that favor individualism, power inequality, and tolerance for uncertainty. Online social media facilitate global access to cultural products, yet this technological capability does not result in cultural convergence [5,6,47]. Instead, consumption of popular videos in culturally different countries appears to be constrained by cultural values.

These findings contrast with studies showing that shared language, common economic system, and geographical proximity are associated with cross-cultural consumption of tangible products [2,7] and flows of information [4,5,8]. The difference with previous results may reflect fewer linguistic, economic, and geographic constraints on video consumption, as well as less need for active interaction with people of different cultures compared to exchanges of e-mail or Tweets, making it easier and more comfortable for YouTube users to encounter and enjoy videos from diverse cultures.

Our findings have implications for the recent upsurge of nationalist movements opposing open borders and free trade. On the one hand, "contact theory" [48] and "soft power" research [49] suggest the possibility that cross-cultural exposure could promote cultural innovation and mutual understanding. On the other hand, cultural openness may erode a country's unique cultural identity, leading to a nationalist backlash.

Additionally, this study has substantial implications for the distinction between cultural betweenness and closeness in cross-cultural experience. Bail [37] highlighted the implications of cultural betweenness as an indicator of cultural bridges – network positions that can "achieve an optimal blend of novelty and familiarity." However, cultural closeness has not received theoretical attention or empirical inquiry. Our findings show that cultural closeness can be an indicator of multicultural identity: countries with low closeness (e.g., Kenya) have a narrow range of video preferences that forms a cultural niche, possibly associated with national identity. In contrast, countries with high closeness (e.g., Canada) have a wide range of video preferences that spans cultural niches and might be associated with a multicultural national identity.

An important limitation of this study is that the units of analysis are countries, not individual users. This poses the possibility that the results we report are susceptible to the ecological fallacy. For example, in some cases, we found that there are significant overlaps in popular video consumption between countries of migration destination and origin. It is possible that individual members of each immigrant group have parochial cultural preferences, but because the groups differ in their preferences, the country appears to be culturally open. We therefore tested for the spurious effects of migration and found significant correlations between cultural openness and a country's degree in the international migration network, where edges correspond to migrations from the country of origin to the country of destination, derived from 2015 UN migration stock data ($r = 0.33, p < .05$ for cultural betweenness; $r = 0.44, p < .001$ for cultural closeness; models including this migration degree as an additional independent variable show identical results with original models, but individualism is no longer significant; more details are provided in S2 Text). Individual user data is needed so that this possibility might be tested more fully in future research. Future research with individual data might also explore possible associations between cultural openness and the incidence of cultural "omnivores" in the population [50].

## Conclusion

Our study makes two important contributions. First, we found that cultural values are significantly associated with the cultural openness of a country, as measured by the consumption of YouTube videos that are popular across diverse cultures. Moreover, this association with cultural values appears to account for effects that previous research has attributed to non-cultural factors. Second, we provide a new angle from which to view the cultural proximity hypothesis in the era of social media on the globalized Web.

25

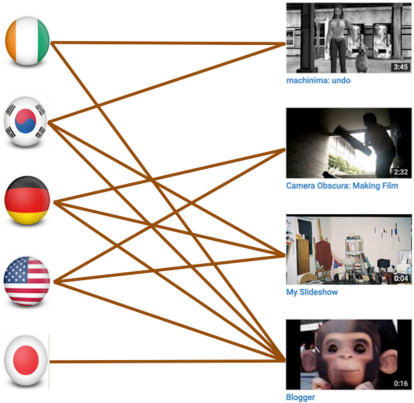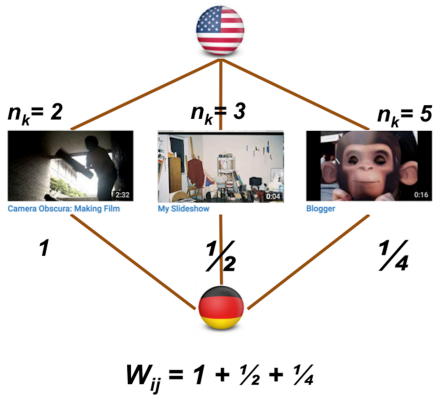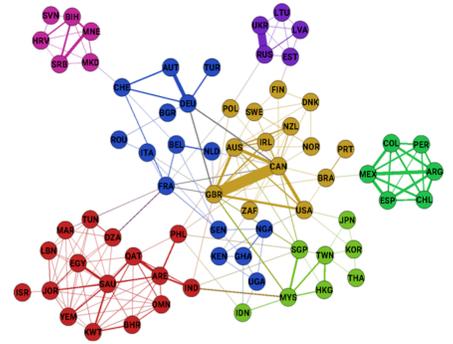

# Supporting information

**S1 Table. Country names included and excluded for analyses.**

**S1 Text. OLS regression analysis using a composite measure of cultural openness.**

**S2 Text. OLS regression analyses including a country's degree in the international migration network as a control variable.**

# Country names included and excluded for analyses.

| | Countries included in the regression | | |
|---|---|---|---|
| ARE | United Arab Emirates | KEN | Kenya |
| ARG | Argentina | KOR | Korea, Republic of |
| AUS | Australia | KWT | Kuwait |
| AUT | Austria | LBN | Lebanon |
| BEL | Belgium | LTU | Lithuania |
| BGR | Bulgaria | LVA | Latvia |
| BRA | Brazil | MAR | Morocco |
| CAN | Canada | MEX | Mexico |
| CHE | Switzerland | MYS | Malaysia |
| CHL | Chile | NGA | Nigeria |
| COL | Colombia | NLD | Netherlands |
| DEU | Germany | NOR | Norway |
| DNK | Denmark | NZL | New Zealand |
| EGY | Egypt | PER | Peru |
| ESP | Spain | PHL | Philippines |
| EST | Estonia | POL | Poland |
| FIN | Finland | PRT | Portugal |
| FRA | France | RUS | Russian Federation |
| GBR | United Kingdom | SAU | Saudi Arabia |
| GHA | Ghana | SEN | Senegal |
| HKG | Hong Kong | SGP | Singapore |
| HRV | Croatia | SRB | Serbia |
| IDN | Indonesia | SVN | Slovenia |
| IND | India | SWE | Sweden |
| IRL | Ireland | THA | Thailand |
| ISR | Israel | TUR | Turkey |
| ITA | Italy | TWN | Taiwan, Republic of China |
| JOR | Jordan | USA | United States of America |
| JPN | Japan | ZAF | South Africa |
| | Countries excluded in the regression | | |
| BHR | Bahrain | OMN | Oman |
| BIH | Bosnia and Herzegovina | QAT | Qatar |
| CZE | Czech Republic | ROU | Romania |
| DZA | Algeria | SVK | Slovakia |
| GRC | Greece | TUN | Tunisia |
| HUN | Hungary | UGA | Uganda |
| MKD | Macedonia, Republic of | UKR | Ukraine |
| MNE | Montenegro | YEM | Yemen |

# OLS regression analysis using a composite measure of cultural openness.

We measured country C's "cultural openness" using a composite measure that is similar to Rao-Stirling diversity, based on the pairwise overlap in popular videos between C and other countries. Conceptually, cultural openness has three dimensions: (1) *breadth*, indicating the number of countries with which C overlaps on at least one video; (2) *spread*, indicating a uniform distribution in the number of overlapping videos with each of those countries; and (3) *heterogeneity*, which measures the extent to which pairwise embeddedness among countries that overlaps popular videos with C. C's cultural openness is the product of breadth, spread, and heterogeneity, such that a country that consumes the same number of videos (*spread*) in common with many other countries (*breadth*) in culturally different clusters (*heterogeneity*) will have a high openness score, whereas a country that disproportionally consumes videos that are popular in a few countries in culturally similar clusters of countries will have a low diversity score.

Operationally, the openness of country $C$ is the weighted product:

$$\sum_{i,j \in N(C)} w(C,i) \times w(C,j) \times d(i,j)$$

where $N(C)$ denotes *breadth* as the list of countries with at least one popular video co-listed with $C$. In this formulation, $w(C,i)$ is the normalized fraction of country $C$'s overlapping videos with another country $i$. The distance $d(i, j)$ indicates Jaccard pairwise dissimilarity between countries *i* and *j* regarding consumption heterogeneity, the overlap of popular video co-listing countries (i.e., 1 − (*N*(*i*) ∩ *N*(*j*))/(*N*(*i*) U *N*(*j*))).

**Table 1. OLS regression model of cultural openness among 58 countries.**

|  | Full model | Non-culture Model | Culture Model |
|---|---|---|---|
| **Intercept** | 0.317 | 0.496*** | 0.411** |
|  | (0.161) | (0.076) | (0.138) |
| **Non-cultural factors** |  |  |  |
| GDP per capita (log-transformed) | -0.117 | -0.191 |  |
|  | (0.206) | (0.237) |  |
| Language eigenvector centrality | 0.003 | 0.171* |  |
|  | (0.08) | (0.081) |  |
| Number of Internet users | 0.258 | 0.296 |  |
|  | (0.225) | (0.252) |  |
| **Cultural values** |  |  |  |
| Individualism (IDV) | 0.295* |  | 0.356** |
|  | (0.130) |  | (0.116) |
| Uncertainty avoidance (UAI) | -0.449*** |  | -0.441*** |
|  | (0.114) |  | (0.106) |
| Power distance (PDI) | 0.450** |  | 0.422** |
|  | (0.146) |  | (0.138) |
| Masculinity (MAS) | 0.226 |  | 0.188 |
|  | (0.127) |  | (0.116) |
| **Sample size (number of countries)** | 58 | 58 | 58 |
| **Model-fit indices** |  |  |  |
| $R^2$ | 0.417 | 0.103 | 0.396 |
| Adjusted $R^2$ | 0.336 | 0.053 | 0.350 |

Note: * $p < .05$, ** $p < .01$, *** $p < .001$. Unstandardized coefficients are reported with standard errors in parentheses. In order to compare coefficients, variables included in the analyses were rescaled to the unit interval, meaning that the minimum value is 0 and the maximum value is 1.

We tested the model for heteroscedasticity and multicollinearity and found heteroscedasticity is not present using non-constant variance score test. Also, the variance inflation factor on each variable of the full model is smaller than 2 except for GDP per capita (2.46) and Internet diffusion (2.51) that are strongly correlated with each other but neither contributes significantly to model predictions.

In short, as shown in Table 2 and 3 in the main text, cultural openness is more closely associated with cultural values than the non-cultural factors.

# OLS regression analyses including a country's degree in the international migration network as a control variable.

As mentioned in the main text, we check the robustness of our results by adding degrees in the migration network as a control variable:

**Table 1. OLS regression model of cultural betweenness among 58 countries.**

|  | Full model | Non-culture Model | Culture Model |
|---|---|---|---|
| **Intercept** | -0.455* | -0.005 | -0.339* |
|  | (0.171) | (0.073) | (0.147) |
| **Non-cultural factors** |  |  |  |
| GDP per capita (log-transformed) | 0.013 | 0.003 |  |
|  | (0.223) | (0.233) |  |
| Language eigenvector centrality | 0.089 | 0.147 |  |
|  | (0.084) | (0.078) |  |
| Number of Internet users | 0.091 | 0.013 |  |
|  | (0.240) | (0.241) |  |
| Degree in Migration Network | 0.193 | 0.184 |  |
|  | (0.119) | (0.101) |  |
| **Cultural values** |  |  |  |
| Individualism (IDV) | 0.240 |  | 0.426** |
|  | (0.167) |  | (0.123) |
| Uncertainty avoidance (UAI) | -0.068 |  | -0.061 |
|  | (0.123) |  | (0.113) |
| Power distance (PDI) | 0.439** |  | 0.373* |
|  | (0.155) |  | (0.147) |
| Masculinity (MAS) | 0.222 |  | 0.250* |
|  | (0.134) |  | (0.124) |
| **Sample size (number of countries)** | 58 | 58 | 58 |
| **Model-fit indices** |  |  |  |
| $R^2$ | 0.417 | 0.164 | 0.396 |
| Adjusted $R^2$ | 0.336 | 0.099 | 0.350 |

Note: * $p < .05$, ** $p < .01$, *** $p < .001$. Unstandardized coefficients are reported with standard errors in parentheses. In order to compare coefficients, variables included in the analyses were rescaled to the unit interval, meaning that the minimum value is 0 and the maximum value is 1.

**Table 2. OLS regression model of cultural closeness among 58 countries.**

|  | Full model | Non-culture Model | Culture Model |
|---|---|---|---|
| **Intercept** | 0.106 | 0.353*** | 0.285** |
|  | (0.102) | (0.051) | (0.100) |
| **Non-cultural factors** |  |  |  |
| GDP per capita (log-transformed) | 0.122 | 0.116 |  |
|  | (0.133) | (0.163) |  |
| Language eigenvector centrality | 0.072 | 0.170** |  |
|  | (0.050) | (0.055) |  |
| Number of Internet users | 0.082 | -0.004 |  |
|  | (0.143) | (0.168) |  |
| Degree in Migration Network | 0.209** | 0.152* |  |
|  | (0.071) | (0.071) |  |
| **Cultural values** |  |  |  |
| Individualism (IDV) | 0.104 |  | 0.328*** |
|  | (0.100) |  | (0.084) |
| Uncertainty avoidance (UAI) | -0.269*** |  | -0.237** |
|  | (0.074) |  | (0.077) |
| Power distance (PDI) | 0.344*** |  | 0.236* |
|  | (0.093) |  | (0.100) |
| Masculinity (MAS) | 0.254** |  | 0.270** |
|  | (0.080) |  | (0.084) |
| **Sample size (number of countries)** | 58 | 58 | 58 |
| **Model-fit indices** |  |  |  |
| $R^2$ | 0.618 | 0.342 | 0.452 |
| Adjusted $R^2$ | 0.555 | 0.291 | 0.411 |

Note: * $p < .05$, ** $p < .01$, *** $p < .001$. Unstandardized coefficients are reported with standard errors in parentheses. In order to compare coefficients, variables included in the analyses were rescaled to the unit interval, meaning that the minimum value is 0 and the maximum value is 1.

**Table 3. OLS regression model of cultural openness (the composite measure) among 58 countries.**

|  | Full model | Non-culture Model | Culture Model |
|---|---|---|---|
| **Intercept** | 0.282 | 0.488 | 0.363* |
|  | (0.160) | (0.077) | (0.139) |
| **Non-cultural factors** |  |  |  |
| GDP per capita (log-transformed) | -0.201 | -0.262 |  |
|  | (0.208) | (0.247) |  |
| Language eigenvector centrality | 0.002 | 0.163 |  |
|  | (0.079) | (0.083) |  |
| Number of Internet users | 0.305 | 0.312 |  |
|  | (0.225) | (0.255) |  |
| Degree in Migration Network | 0.115 | 0.089 |  |
|  | (0.125) | (0.107) |  |
| **Cultural values** |  |  |  |
| Individualism (IDV) | 0.269 |  | 0.403** |
|  | (0.156) |  | (0.117) |
| Uncertainty avoidance (UAI) | -0.466*** |  | -0.439*** |
|  | (0.115) |  | (0.104) |
| Power distance (PDI) | 0.485** |  | 0.464** |
|  | (0.145) |  | (0.139) |
| Masculinity (MAS) | 0.230 |  | 0.186 |
|  | (0.125) |  | (0.115) |
| **Sample size (number of countries)** | 58 | 58 | 58 |
| **Model-fit indices** |  |  |  |
| $R^2$ | 0.454 | 0.120 | 0.423 |
| Adjusted $R^2$ | 0.363 | 0.052 | 0.378 |

Note: * $p < .05$, ** $p < .01$, *** $p < .001$. Unstandardized coefficients are reported with standard errors in parentheses. In order to compare coefficients, variables included in the analyses were rescaled to the unit interval, meaning that the minimum value is 0 and the maximum value is 1.